\documentclass[preprintnumbers,amsmath,amssymb,floatfix]{revtex4}
\usepackage{graphicx}
\usepackage{dcolumn}
\usepackage{bm}

\def\be{\begin{equation}}
\def\ee{\end{equation}}
\def\ba{\begin{array}}
\def\ea{\end{array}}
\def\bea{\begin{eqnarray}}
\def\eea{\end{eqnarray}}

\begin{document}

\title{Entropy and light cluster production in heavy-ion collisions at intermediate energies}
\author{Yogesh K. Vermani}

\author{Rajeev K. Puri}
\email{rkpuri@pu.ac.in} \affiliation{Department of Physics, Panjab University, Chandigarh -160 014, India}

\date{\today}

\begin{abstract}
The entropy production in medium energy heavy-ion collisions is
analyzed in terms of ratio of deuteronlike to protonlike clusters
($d_{like}/p_{like}$) using \emph{quantum molecular dynamics}
(QMD) model. The yield ratios of deuteronlike-to-protonlike
clusters calculated as a function of participant proton
multiplicity closely agree with experimental trends. Our model
predictions indicate that full thermodynamical equilibrium may not
be there even for the central geometry. The apparent entropy
extracted from the yield ratios of deuteronlike-to-protonlike
clusters, however, reflects the universality characteristics
\emph{i.e.} it is governed by the volume of reaction independent
of the target-projectile combination. Our calculations for
apparent entropy produced in central collisions of Ca+Ca and Nb+Nb
at different bombarding energies are in good agreement with $4\pi$
Plastic Ball data.
\end{abstract}

\maketitle

\section{Introduction}

Various experimental and theoretical studies in the recent past
have indicated a clear demarkation of colliding matter into
participant and spectator matter especially at high incident
energies \cite{kunde, schaun}. This is characterized by the
formation of hot \& dense fireball. In addition, the size of the
participant volume or `source size' is also linked with the
emission of composite particles and ultimately with the mechanism
of the production of entropy \cite{sim, mish, rem}. The highly
dense fireball formed for a very short duration ($<10^{-22}s$) can
be of importance for probing the properties of condensed nuclear
matter. The formation of the fireball is reported to be affected
by many factors such as beam energy, overlapping volume, as well
as density reached in a reaction. Earlier experimental studies
using the Plastic Ball/Wall detector \cite{doss, doss88} showed
that the deuteron-to-proton yield ratio varies with the impact
parameter of the reaction indicating a strong dependence on the
participant volume. Such behavior is believed to be resulted from
the coalescence production mechanism \cite{doss, doss88}.
Experimentally, the study of $d/p$ ratios obtained in asymmetric
collisions of p+Kr, O+Kr, and Ne+Ar also indicates the importance
of coalescence mechanism \cite{fokin}. In this study, it was
observed that $d/p$ ratio increases with the beam energy. The
baryonic entropy, however couldn't be directly measured from these
ratios, since total yield is required from the source. The
measurement of the yield ratios and baryonic entropy produced are,
therefore, very promising candidates for estimating the fireball
produced in the hot \& dense nuclear matter.

As a matter of fact, the pion production also contributes
significantly \cite{stach98, andro, ahle, bart} towards the
entropy generation at SPS \cite{stach99, fanas} and higher
energies. At SPS energies ($\sim$160 AGeV fixed target), the pion
number increases with beam energy to about ten times the number of
the nucleons \cite{fanas}. There, an increase in the entropy
production was observed with beam energy as one moves from AGS
energies towards SPS and higher energies \cite{andro}. This
enhancement may also be conjectured as manifestation of the change
in the collision dynamics at such high energy \cite{marek95,
marek95a}. For the incident energy range considered (400-1050
AMeV) in the present work, the inclusion of the pion production is
not going to affect the entropy production appreciably \cite{ahle,
bart, klay}. For instance, at SIS energies (upto 2 AGeV), the
total number of pions is only 10 \% of the nucleons \cite{hong}.

In the present work, we aim to estimate the baryonic entropy
$S_{N}$ generated in the fireball via the yields of free particles
and light clusters (\emph{i.e.} protons, neutrons, deuteron,
tritium, helium-3, and $\alpha$-particles). We made a systematic
study of symmetric heavy-ion (HI) collisions in the incident
energy range 400-1050 AMeV within dynamical model namely,
\emph{quantum molecular dynamics} (QMD) model \cite{aich}. We
shall also compare our results with experimental data taken with
Plastic Ball/Wall detector. The dynamical approach such as QMD
model provides an useful platform to estimate the yield of
composite particles in terms of \emph{n-n} correlations from the
start when the colliding nuclei are well separated to the final
state where matter is cold and fragmented. As reported in Ref.
\cite{jk}, the production of light charged particles carries vital
information about the stopping and equilibration of nuclear matter
during a collision.

\section {The Quantum Molecular Dynamics Model}

The quantum molecular dynamics (QMD) model is a \emph{n}-body
transport theory that describes HI reactions in intermediate
energy regime (20 MeV/nucleon $\leq E_{lab}\leq$ 2 GeV/nucleon) on
event by event basis. The two essential ingredients of the model
are nucleon-nucleon (\emph{n-n}) potential and stochastic
scattering. Here each nucleon follows the trajectory according to
the classical equations of motion \cite{aich}:
\begin{equation}
\dot{{\bf p}_i}=-\frac{\partial \langle {\cal H} \rangle}{\partial
{\bf r}_i}, ~ \dot{{\bf r}_i}=\frac{\partial \langle {\cal H}
\rangle}{\partial {{\bf p}_i}}.
\end{equation}
The expectation value of the total Hamiltonian $\langle {\cal H}
\rangle$ consists of kinetic and potential energies terms as:
\begin{eqnarray}
\langle {\cal H} \rangle &=& \langle {\cal T} \rangle + \langle {\cal V} \rangle  \nonumber \\
&=& \sum_{i=1}^{A}{\frac{{\bf p}_i^2}{2m_i}} + \sum_{i=1}^{A}
(V_{i}^{Skyrme}+V_{i}^{Yuk}+V_{i}^{Coul} +V_{i}^{MDI}). \label{H}
\end{eqnarray}

The $V_{i}^{Skyrme}$, $V_{i}^{Yuk}$, $V_{i}^{Coul}$ and
$V_{i}^{MDI}$ in Eq.(\ref{H}) are, respectively, the local Skyrme,
long range Yukawa, an effective charge Coulomb and momentum
dependent parts of the interaction \cite{aich}. The potential part
without momentum dependent interactions (MDI) is called static
equation of state (EoS). For the present study, we employ a soft
equation of state represented by incompressibility
$\kappa=200~MeV$ and energy dependent \emph{n-n} cross section
\cite{aich}. The equation of state with MDI is labeled as `SM'
whereas without MDI, it is labeled as `S'. Since many studies
reveal the nuclear matter to be softer \cite{hdad, ross}, we also
prefer the soft EoS.

The entropy information may be obtained from a classical charge
symmetric \emph{gas} of nucleons and deuterons in thermal and
chemical equilibrium using the relation suggested by Kapusta
\emph{et al.} \cite{sim, cse}:
\begin{equation}
S_{N}= 3.945-\ell nR_{dp}, \label{rd}
\end{equation}
where $R_{dp}$ is the ratio of deuterons to protons established
during early stages of the fireball. One source of error arises
due to the neglect of other light composite particles \emph{viz.}
t, $^{3}He$ and $\alpha$. Bertsch and Cugnon \cite{bert} proposed
to take into account these lighter fragments as well generalizing
Eq.(\ref{rd}) as:
\begin{equation}
S_{N}= 3.945-\ell n~x, \label{rdp}
\end{equation}
where
\begin{eqnarray}
x&=& d_{like}/p_{like} \nonumber \\
&=&
\frac{d+\frac{3}{2}t+\frac{3}{2}~^{3}He+3\alpha}{p+d+t+2~^{3}He+2
\alpha}.\label{dlike}
\end{eqnarray}
\noindent As seen, quantity `$x$' measures the yield ratio of
deuteron-like ($d_{like}$) to proton-like ($p_{like}$) fragments.
It has been well established from experiments that highest proton
multiplicity accounts for most of the charges in HI system, thus
leaving no room for heavier clusters. Following Ref. \cite{peil},
we define the yield ratio of deuteron-like ($d_{like}$) to
proton-like ($p_{like}$) clusters in the following way:
\begin{equation}
\tilde{R}_{dp}=\frac{Y(A=2)+\frac{3}{2}Y(A=3)+3Y(A=4)}{N_{p}},
\end{equation}
where Y(A) stands for the number of fragments with mass `A' in one
event. Analogous to experimental results, we calculate the total
participant multiplicity $N_{p}$ as:
\begin{eqnarray}
N_{p}=&\frac{Z_{P}+Z_{T}}{A_{P}+A_{T}}[Y(A=1)+2Y(A=2)  \nonumber \\
      & +3Y(A=3)+4Y(A=4)],
\end{eqnarray}
where $Z_{P}+Z_{T}$ and $A_{P}+A_{T}$ define the total charge and
mass of the colliding system. This procedure allows us to
calculate the entropy produced in a reaction. We have calculated
these ratios within QMD model for unfiltered events, employing
\emph{minimum spanning tree} (MST) procedure of clusterization
\cite{mst}. In this clusterization approach, we assume that two
nucleons sharing the same cluster if their centroids are closer
than a spatial radius $ r_{C}=|{\bf r}_{\alpha}-{\bf r}_{\beta}|$.
One generally choose the radius $r_{C}=4$ fm. It may be mentioned
that some other sophisticated algorithms have also been developed
which are based on relative momentum of nucleons \cite{jkd},
fragments' binding energy minimization \cite{yugs}, and
\emph{backtracing} procedure \cite{bond} to study the fireball and
spectator matter physics.

\section{Results and Discussion}

Since it is well established that the production of light charged
particles and clusters, and ultimately the entropy is related to
the fireball, it is of interest to see their phase space. In Fig.
1, we display the space (X-Z) of $^{93}Nb+^{93}Nb$ collision at
650 AMeV and at a reduced impact parameter $b/b_{max}$=0.6. The
chosen time represents the freeze out time when density reaches
asymptotic value. First of all, consistent with earlier attempts,
we see that heavier fragments belong to the residue of either
projectile or target, whereas lighter clusters like free nucleons
and light charged particles LCPs [$2\leq A \leq 4$] are produced
due to the coalescence and emerge from the mid-rapidity region. As
shown in Ref. \cite{jk}, these light charged particles carry vital
information about the stopping as well as thermalization of the
nuclear matter, therefore, are also good candidates for the study
of production of entropy in HI reactions. A very little influence
can be seen of momentum dependent interactions.

Next we study the final state composite particle yield ratios X/p
for the soft (S) and soft momentum dependent (SM) interactions.
This is shown in Fig. 2 for the collisions of $^{93}Nb+^{93}Nb$ at
incident energy of 650 AMeV as a function of impact parameter.
Here `X' stands for A=2, 3 and 4 clusters.
On can see that curves for different X/p ratios exhibit similar trends: \\
\noindent (i) The X/p ratio decreases with impact parameter
(alternately, increases with $N_P$) indicating more production in
central collisions compared to peripheral collisions. As shown by
many authors \cite{rem, doss, peil}, $N_{P}$ remains same for
nearly central collisions and decrease sharply for semi-central
collisions and peripheral collisions. \\
(ii) For central impact parameters (or, higher $N_P$ values), X/p
ratios reach an asymptotic value indicating that for central events,
small variation in impact parameter does not give different results. \\
(iii) Role of momentum dependent interactions is nearly marginal
justifying the earlier attempts \cite{aich87} and use of soft equation of state. \\

Since entropy production is mostly measured for central
collisions, the use of momentum dependent interactions will not
give different results compared to static soft equation of state.
These different yield ratios implied that one also obtains
different behavior of density reached, collision rate and
multiplicity of various light mass fragments for S and SM
interactions.

In Fig. 3, we extend the above study by including the ratio of
deuteron-like ($d_{like}$) to proton-like ($/p_{like}$) clusters.
The calculations for $d_{like}/p_{like}$ are made using soft (S)
equation of state for the collisions of $^{40}Ca+^{40}Ca$ (at 400
and 1050 AMeV) and $^{93}Nb+^{93}Nb$ (at 400 and 650 AMeV) as a
function of participant proton multiplicity. The results of the
Plastic Ball experiments \cite{doss} are also displayed  for
comparison. The Plastic ball data takes into account the overlap
region for the yield of deuteronlike and protonlike clusters,
while our ratios are calculated for the unfiltered events using
MST procedure. The yield ratios are calculated typically after 40
fm/c, when average nucleonic density saturates and \emph{n-n}
collisions practically cease. At this time, yields of  composite
particles is well established and may be compared with
experimental data. One clearly sees that our model describes well
the functional form of experimental $d_{like}/p_{like}$ ratio
which is found to increase with $N_P$ (or centrality of the
collision) and saturates at higher multiplicity end. At low $N_P$,
there is a large drop in the yield ratio for model calculations as
also found for individual cluster-to-proton (X/p) ratios (See
Fig.2). These trends are closely related with nuclear matter
stopping and flow effect in the formation of hot and dense
fireball. Recently, Dhawan \emph{et al.} \cite{jk} studied impact
parameter dependence of light charged particles (LCPs) yield and
anisotropy ratio. It was found that LCPs production was maximum at
central collisions where maximum stopping of nuclear matter is
also achieved. Thus, production of light clusters can act an
indicator of global stopping achieved in the nuclear matter.
Interestingly, $d_{like}/p_{like}$ ratio calculated using
dynamical approach is in good agreement with experimental data.
This shows that one can reliably explore the applicability of
dynamical approach such as QMD model to further investigate the
formation of fireball at intermediate energies.

It should be emphasized that model of Kapusta \cite{sim,mish} is
highly idealized one that relies on the assumption of thermal and
chemical equilibrium in the system, for extraction of $S_{N}$
using Eqs.~(\ref{rd}) and (\ref{rdp}). The heavy-ion collisions at
higher incident energies evolve from an initial non-equilibrium
momentum distribution to a thermalized distribution. In a
heavy-ion system, full equilibrium is not, however, always
possible even at central geometry \cite{khoa}. At large impact
parameters, the momentum distribution of nucleons further deviates
from the degree of full equilibrium \cite{khoa}. It is worthwhile
to study the equilibration of the nuclear matter produced at the
time of fireball formation and role of secondary de-excitation of
heavier clusters on proton-like and deuteron-like abundances
during the expansion stage. The anisotropy ratio, which is also a
measure of the degree of equilibration reached in a heavy-ion
reaction, is defined as \cite{jk, khoa}:
\begin{equation}
\langle R_{A} \rangle = \frac{ \sqrt{ \langle p_{x}^{2} \rangle} +
\sqrt{\langle p_{y}^{2} \rangle}}{2\sqrt{\langle p_{z}^{2}\rangle
}}. \label{ar}
\end{equation}
The full equilibrium corresponds to $\langle R_{A} \rangle$ value
close to unity. In Fig. 4(a), we plot the time evolution of
anisotropy ratio $R_{A}$ for the reactions of Nb+Nb at incident
energies of 400 and 650 AMeV, respectively. The shaded area
corresponds to the time zone for highly excited nuclear matter
which is followed by the decompression phase. Beyond this region,
the nucleon density saturates and hard \emph{n-n} collisions cease
almost. One can see that $\langle R_{A} \rangle$ values approach
close to asymptotic value just after the compression (t$\sim$
40-45 fm/c) irrespective of the incident energy chosen. The
$\langle R_{A} \rangle$ values at both incident energies are still
less than 1. It clearly shows that the HI system has still not
achieved full equilibrium at the expansion stage of the collision.
It also indicates that \emph{n-n} collisions occurring at later
times don't alter the momentum space of nucleons significantly. In
the lower panels (b)-(c), we display the yields of $p_{like}$ and
$d_{like}$ clusters. One can see that in low-density phase, the
secondary emission from heavier clusters tend to populate the
abundances of deuteron-like and proton-like clusters slightly.

Since full thermodynamical equilibrium is not visible for the
participant zone, we shall estimate \emph{apparent} entropy
`$S_{app}$' produced in the HI reactions using Eq.(\ref{rdp}). We
display in Fig. 5, the apparent entropy $S_{app}$ as a function of
participant proton multiplicity $N_{P}$ for the reactions of
$Ca+Ca$, $Nb+Nb$, and $Au+Au$ at incident energies of 400 and 650
AMeV. One can clearly see that at a given beam energy, it is the
volume of participant nucleons (that is, $N_{P}$) which governs
the apparent entropy produced rather than the total number of
nucleons in the phase space. This clearly brings out the
participant-spectator picture of HI collisions at relativistic
beam energies. It depicts that participant volume is solely
contributing towards deuteronlike and protonlike cluster
abundances independent of the target-projectile combination. These
results are in agreement with the experimental data and
theoretical approaches. It may be mentioned that apparent entropy
of the fireball would approach realistic baryonic entropy using
Eq.(\ref{rdp}) only when full thermodynamical equilibrium is
achieved for the participant zone.

Finally we display in Fig. 6, the apparent entropy `$S_{app}$'
extracted for the central collisions of $Ca+Ca$ (at 400 and 1050
AMeV) and $Nb+Nb$ (at 400 and 650 AMeV). Also shown here is the
baryonic entropy obtained in Plastic Ball experiment \cite{doss}
using the model of Kapusta. Our model entropy values are in close
agreement with experimental observations \cite{doss, nag}. Nearly
no effect of beam energy is visible in experimental as well as in
model entropies.

\section{Summary}
In this paper, we have modeled the dynamical emission of light
clusters in highly excited nuclear matter and estimated specific
entropy based upon the formalism proposed by Kapusta \emph{et al}.
Our model calculations are done within the framework of
\emph{quantum molecular dynamics} (QMD) model for unfiltered
events. The comparison of model calculations of yield ratios as a
function of participant proton multiplicity $N_{P}$ with
experimental data gives quite encouraging results. From these
results, we conclude that it is the volume of participants in a
reaction system that governs deuteronlike and protonlike cluster
abundances in early interaction stage. The time evolution of
momentum anisotropy ratio shows that full thermodynamical
equilibrium isn't, however, achieved even at later expansion stage
of the heavy-ion collisions. Our model calculations for
\emph{apparent} entropy produced in the central collisions of
Ca+Ca and Nb+Nb indicate close agreement with experimental results
based upon Kapusta's formalism. Nearly no effect of beam energy is
visible in experimental as well as in model entropies.


\newpage
\noindent {\Large \bf Figure Captions} \\

{\bf FIG. 1.} The snapshots of spatial coordinates of nucleons in
X-Z plane for the reaction of Nb(650 AMeV)+Nb at reduced impact
parameter $b/b_{max}$= 0.6. Time taken here corresponds to the
case when \emph{n-n} collisions cease almost just after the
violent phase of the reaction. Different colors depict the free particles, nucleons bound in light charged particles as well as in heavier fragments. \\

{\bf FIG. 2.} The yield ratio of light clusters to protons (X/p)
as a function of impact parameter b using minimum spanning tree
procedure. The term `global' for ratios X/p signifies that
particle yield is calculated taking full
ensemble into account and not the limited region only.\\

{\bf FIG. 3.} The $d_{like}/p_{like}$ ratio as a function of
baryon charge multiplicity $N_{p}$. The model calculations (open
symbols) at the time of freeze out are compared with experimental
data (solid symbols). The results are shown here for the reactions
of Ca+Ca (l.h.s.) and Nb+Nb (r.h.s.). \\

{\bf FIG. 4.} The anisotropy ratio $\langle$R$_{A} \rangle$, and
the yields of deuteron-like ($d_{like}$) and proton-like
($p_{like}$) clusters obtained during the evolution of central
$^{93}Nb+^{93}Nb$ collisions at incident energies of 400 and 650 AMeV. \\

{\bf FIG. 5.} Apparent entropy per nucleon $S_{app}$ as a function
of baryon charge multiplicity $N_{p}$ for the reactions of
$^{40}Ca+^{40}Ca$ (open circles), $^{93}Nb+^{93}Nb$ (half filled
circles), and $^{197}Au+^{197}Au$ (open squares). Calculations
shown here are at incident energies of 400(top) and 650(bottom) AMeV. \\

{\bf FIG. 6.} The relationship between baryonic entropy and beam
energy $E_{lab}$ for the central collisions of $^{40}Ca+^{40}Ca$
and $^{93}Nb+^{93}Nb$. Also shown are entropy values extracted by
the Plastic Ball group \protect\cite{doss} using Kapusta's formalism. \\

\newpage

\begin{figure}[!t]
\centering \vskip -1cm
\includegraphics[scale=0.65] {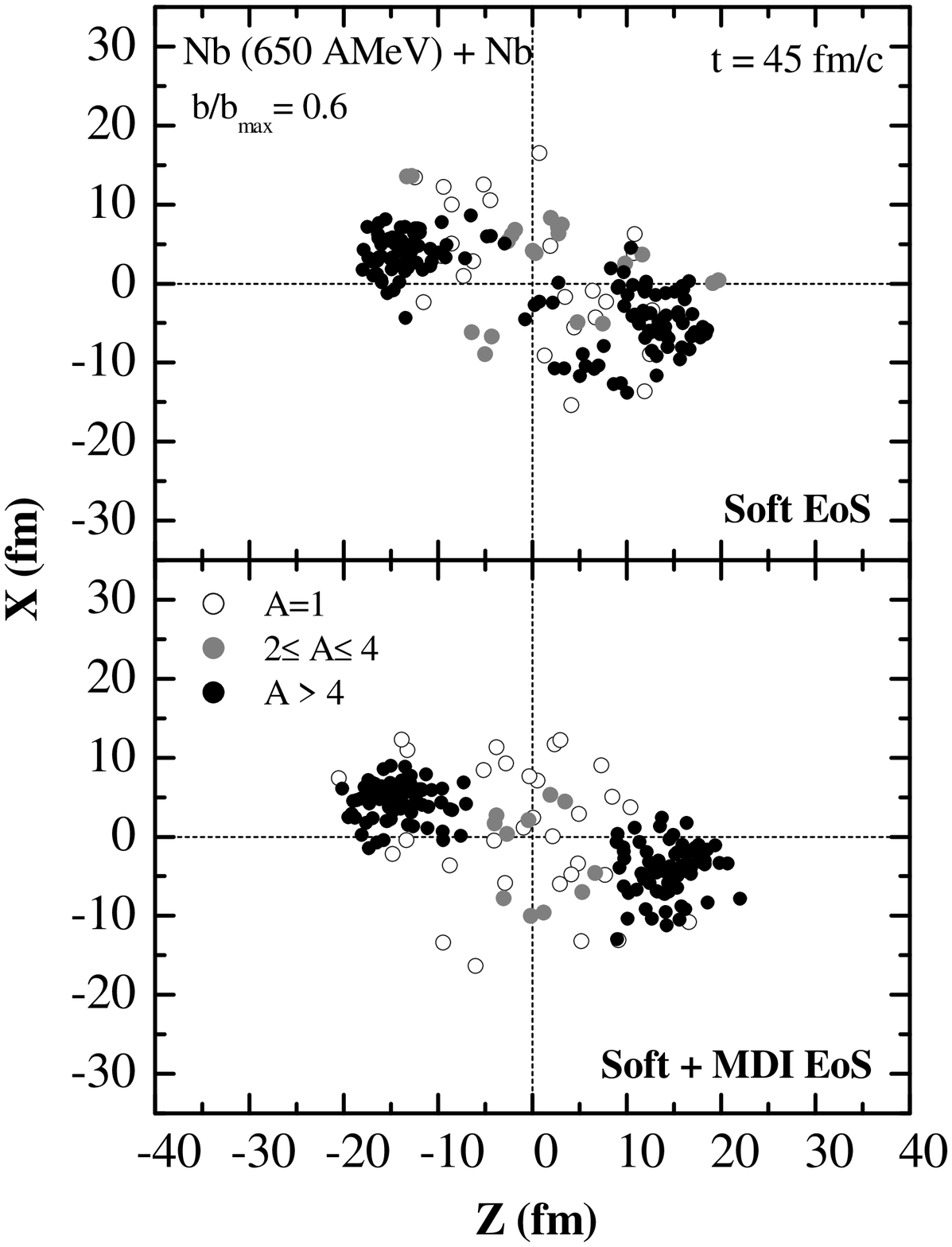}
\caption { }
\end{figure}

\begin{figure}[!t]
\centering \vskip -1cm
\includegraphics[scale=0.65] {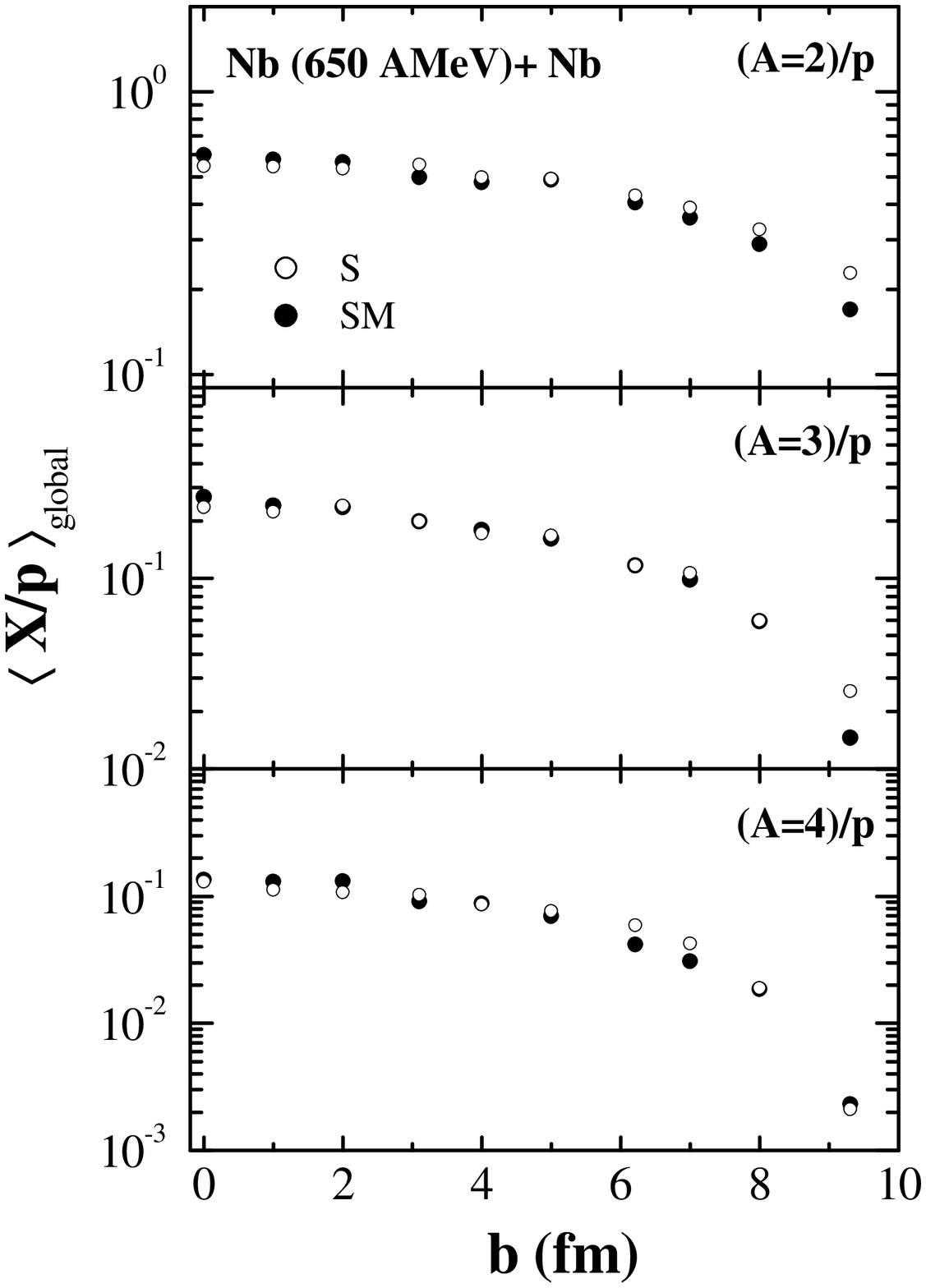}
\caption { }
\end{figure}

\begin{figure}[!t]
\centering \vskip -1cm
\includegraphics[scale=0.65] {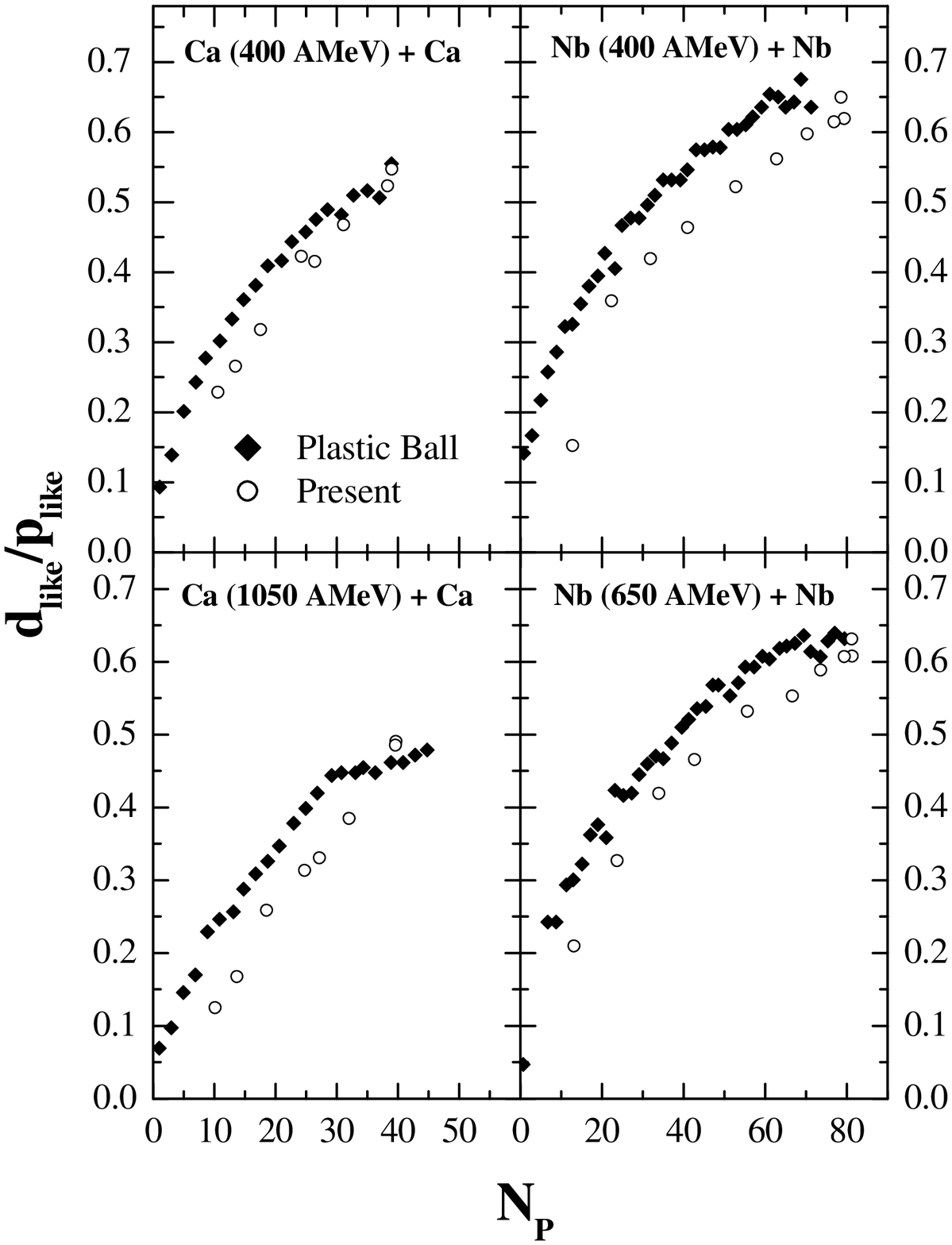}
\caption { }
\end{figure}

\begin{figure}[!t]
\centering \vskip -1cm
\includegraphics[scale=0.65] {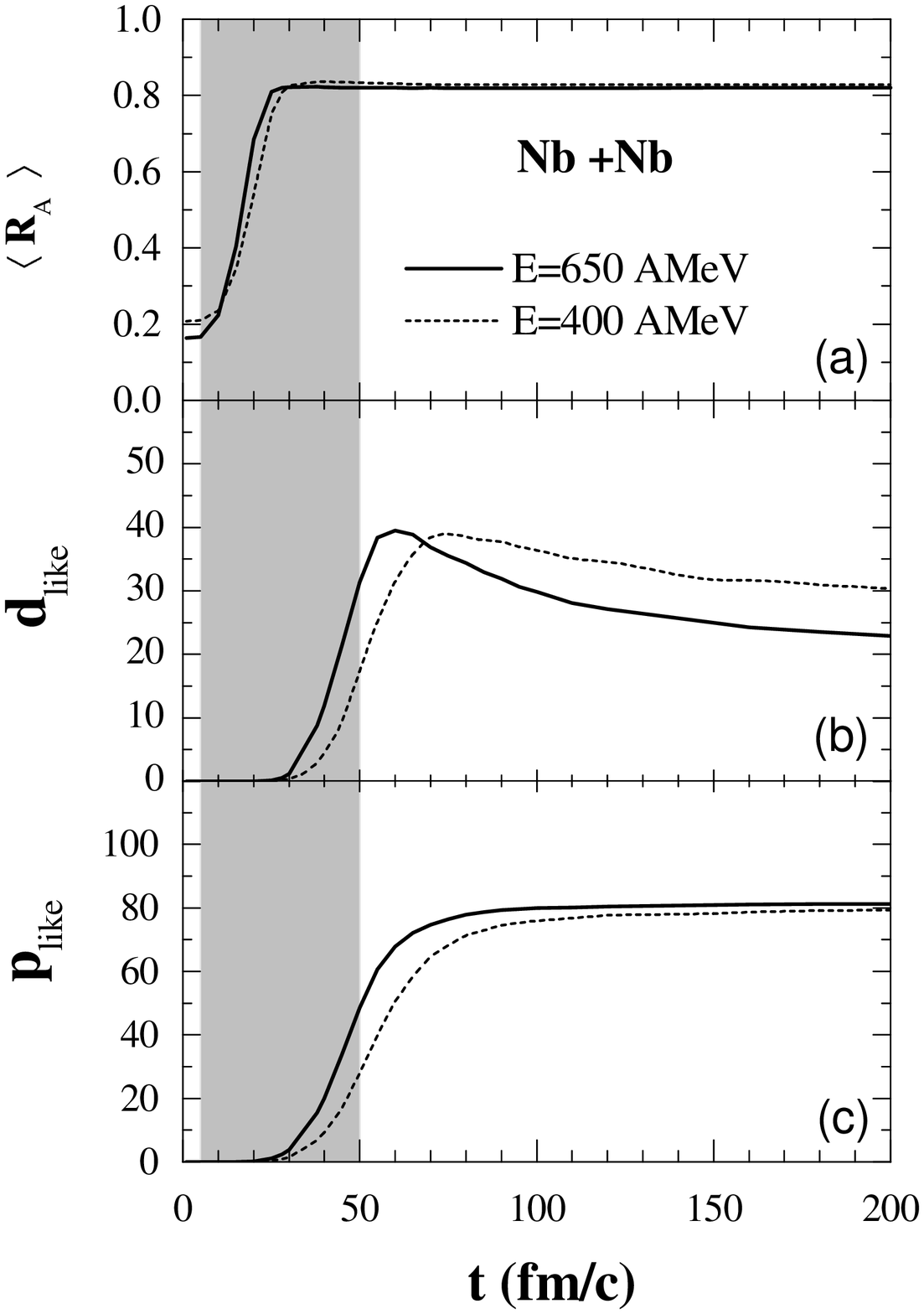}
\caption { }
\end{figure}

\begin{figure}[!t]
\centering \vskip -1cm
\includegraphics[scale=0.65] {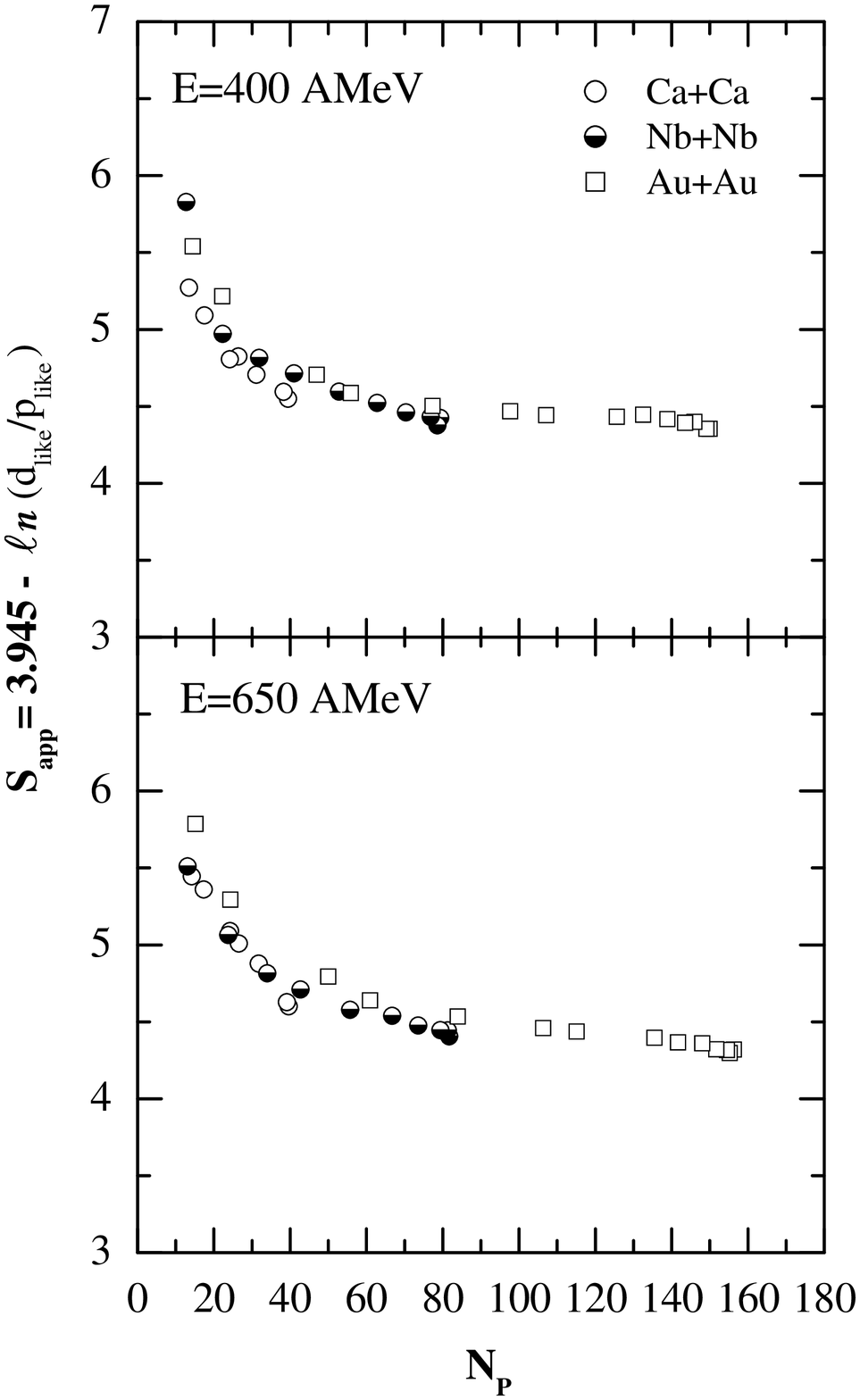}
\caption { }
\end{figure}

\begin{figure}[!t]
\centering \vskip -1cm
\includegraphics[scale=0.65] {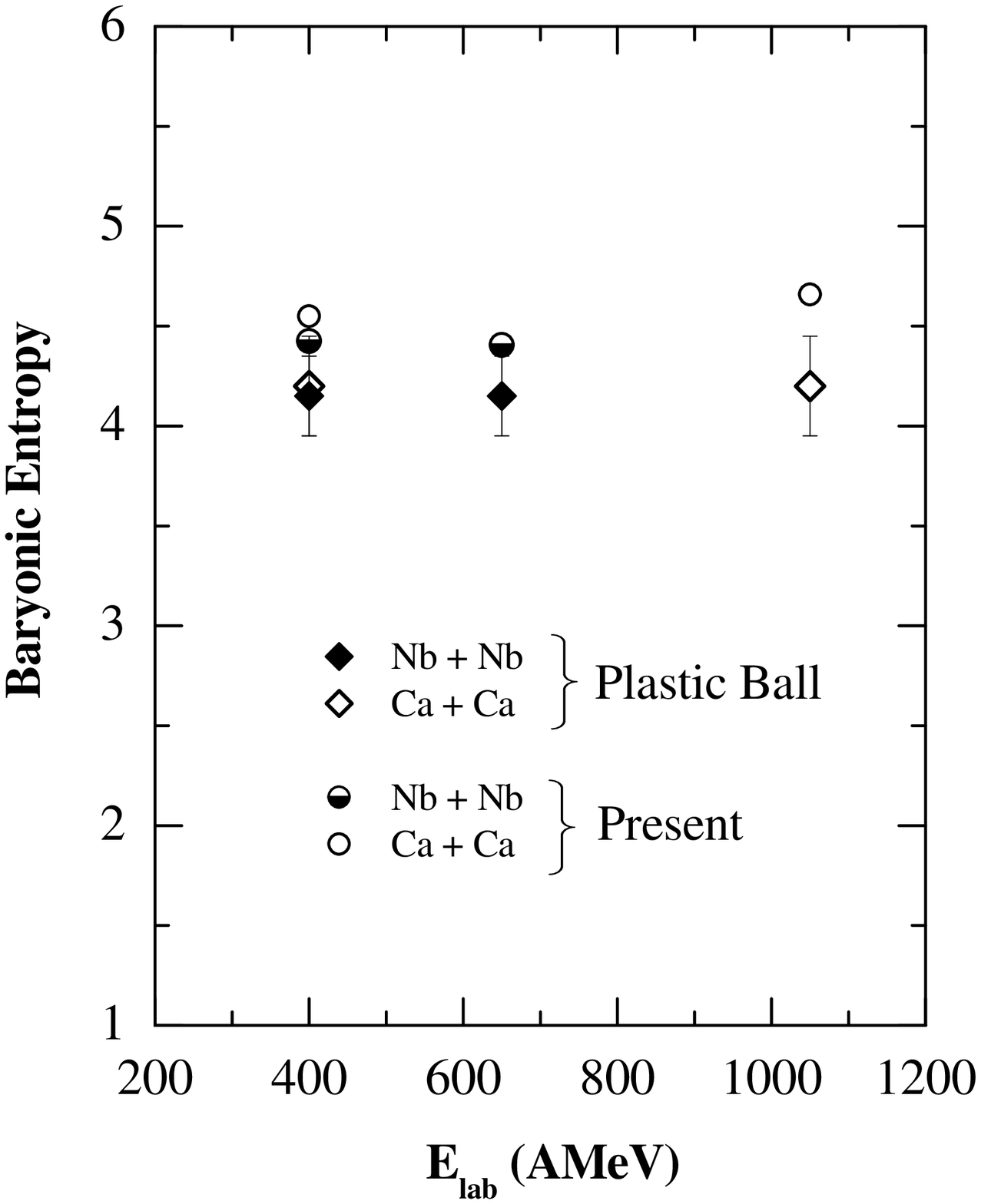}
\caption { }
\end{figure}

\end{document}